\documentclass[conference]{IEEEtran}
\IEEEoverridecommandlockouts

\usepackage{booktabs}
\usepackage{amsmath}
\usepackage{amsthm}
\usepackage{mathrsfs}
\usepackage{mathtools}
\usepackage{subfigure}
\usepackage{graphicx}
\usepackage{latexsym}
\usepackage{subeqnarray}
\usepackage{cases}
\usepackage{amssymb,amsbsy,verbatim,array}
\usepackage{epsfig}
\usepackage{epstopdf}
\usepackage{array}
\usepackage{cite}
\usepackage{color}
\usepackage[ruled,lined,linesnumbered]{algorithm2e}
\usepackage{makecell}
\usepackage{extarrows}
\usepackage{multicol}
\usepackage{multirow}
\usepackage{bm}
\usepackage{mathrsfs}
\usepackage{booktabs}
\usepackage{siunitx}
\usepackage[switch]{lineno}
\usepackage{tabularx,booktabs}
\usepackage[flushleft]{threeparttable}
\usepackage{stfloats}

\usepackage[colorlinks,
            linkcolor=red,
            anchorcolor=blue,
            citecolor=green,
            ]{hyperref}

\begin{document}

\title{Multi-source Scheduling and Resource Allocation for Age-of-Semantic-Importance Optimization in Status Update Systems
\thanks{
This research was funded by the National Natural Science Foundation of China grant number 62171481 and Technology Program of Guangzhou under Grant 202201011577.}
}

 \author{Lunyuan Chen, Jie Gong\\ \textit{Guangdong Key Laboratory of Information Security Technology}\\\textit{School of Computer Science and Engineering, Sun Yat-Sen University, Guangzhou, 510006, China}\\\textit{Email: chenly286@mail2.sysu.edu.cn, gongj26@mail.sysu.edu.cn
}
}

\maketitle

\begin{abstract}
In recent years, semantic communication is progressively emerging as an effective means of facilitating intelligent and context-aware communication. However, current researches seldom simultaneously consider the reliability and timeliness of semantic communication, where scheduling and resource allocation (SRA) plays a crucial role. In contrast, conventional age-based approaches cannot seamlessly extend to semantic communication due to their oversight of semantic importance.
To bridge this gap, we introduce a novel metric: Age of Semantic Importance (AoSI), which adaptly captures both the freshness of information and its semantic importance. Utilizing AoSI, we formulate an average AoSI minimization problem by optimizing multi-source SRA. To address this problem, we proposed a AoSI-aware joint SRA algorithm based on Deep Q-Network (DQN).
Simulation results validate the effectiveness of our proposed method, demonstrating its ability to facilitate timely and reliable semantic communication.
\end{abstract}


\section{Introduction}
\label{sec:1}
Over the past few years, semantic communication has emerged as a pivotal paradigm for future wireless communications. In semantic systems, a source transmits semantic information extracted from original data while minimizing semantic-level loss rather than merely correcting bit-level errors during transmission\cite{Semantic1}. 
The evolution of semantic communication encompasses significant developments in semantic information theory, including semantic rate distortion and deep learning (DL)-based data compression\cite{semantic4}. Moreover, application-centric approaches have also emerged, such as DL-based joint source-channel coding (JSCC), which effectively harness DL to enhance the efficiency and reliability of communication systems\cite{DeepSC,semantic2,semantic3}. Ref.\, \cite{DeepSC} studied the semantic text transmission which uses the Transformer model to enhance system capacity and reduce semantic errors by sentence meaning recovery. The authors in \cite{semantic2} investigated wireless image transmission and minimized the semantic level loss.

To improve the performance of semantic communication, especially in the realm of deep JSCC, researchers have introduced various new metrics and delved deeply into semantic resource allocation\cite{Qin1,Qin2,semanticRA1,semanticRA2}. Ref.~\cite{Qin1} pioneered a novel metric, semantic spectral efficiency and optimized resource allocation for text-based semantic communication. Authors in \cite{Qin2} defined semantic entropy for different tasks and addressed the Quality of Experience (QoE) maximization problem. For image transmission, ref.~\cite{semanticRA2} focused on optimizing compression ratio, power allocation, and resource block assignment to enhance user numbers and image quality. Furthermore, authors in \cite{GuoWLS23} introduced a metric called semantic importance to measure semantic loss in transmission and explored semantic-importance-aware communication schemes using pre-trained language models. Despite these advancements, the aspect of information freshness within the deep JSCC framework remains under-explored.

Age of information (AoI), defined as the time elapsed since the
generation of the latest received update, is a well-known metric for information freshness\cite{AoI0}. It has been extensively studied in the literature (see \cite{AoI1} and references therein). However, it only cares about if the received data is outdated or not compared with generation time, but ignores the accuracy and effectiveness of received information. In fact, the outdated data is still accurate as long as the source's status does not change. To deal with this issue, age of incorrect information (AoII) has been proposed to capture the deteriorating effect the incorrect information\cite{AoII1}. This concept has been adapted for semantic-aware scenarios in \cite{AoII3}, where content-aware sampling policies were implemented based on AoII to simultaneously evaluate the freshness and value of status updates.
However, for application-centric approaches, especially within the deep JSCC framework, there are few studies on the semantic-aware scheduling and resource allocation (SRA) considering both the information freshness and semantic importance.

In this paper, we investigate a semantic-aware multi-source status updates system, involving multiple sources and one edge server. To evaluate the system performance in terms of information freshness and semantic loss, we propose a novel metric named Age of Semantic Importance (AoSI). We formulate an average AoSI minimization problem by optimizing SRA. Specifically, we propose a Deep Q-Network (DQN)-based algorithm for SRA. The main contributions of this paper are summarized as follows:

\begin{itemize}
\item We explore a multi-source status update system where multiple sources transmit semantic information extracted from their status update packets to an edge server through semantic communication, thereby negating the need to send raw data directly.

\item We introduce AoSI as a novel metric to quantify both the freshness and semantic similarity of the transmitted updates. Utilizing this metric, we provide the analysis of the long-term average AoSI of the network.

\item We formulate and address the challenge of minimizing average AoSI through source scheduling and resource allocation, and then proposed a DQN-based algorithm to obtain the SRA policy.

\item Through simulations, we demonstrate the effectiveness of our AoSI-aware joint SRA algorithm, showcasing its superiority over baseline methods.
\end{itemize}
The remaining parts of this paper are arranged as follows. Section \ref{sec:2} describes the system model of the considered semantic-aware multi-source status update system. In section \ref{sec:3}, we introduce the definition of the novel metric AoSI and analyze the long-term average AoSI. Section \ref{sec:4} formulates the average AoSI minimization problem by optimizing source scheduling and resource allocation, and then solves the problem using a DQN-based SRA algorithm. The simulation results and discussions are provided in Section \ref{sec:5}. Finally, we come to the conclusion in Section \ref{sec:6}.

\section{System Model}
\label{sec:2}
\subsection{Network model}
As illustrated in Fig. 1, we consider a semantic-aware multi-source status update system, wherein multiple sources transmit their status packets to an edge server via semantic communication. Let $\mathcal{M} = \{1, 2, \dots, M\}$ represent the source set. Each source is equipped with a semantic transmitter, while the edge server is fitted with a semantic receiver to enhance uplink semantic communication. In this system, each source periodically generates status updates with sampling period $\tau$. At the start time $t_n$ of each sampling period $n, n=1,2,\dots, N$, a source $m$ generates packet $l_{m,n}$ with probability $P_m$. A central controller located at the edge server then determines which source will transmit its status packet to the edge server. The corresponding scheduling decision is denoted as $\mathbf{\alpha}_n = \{\alpha_{1,n}, \alpha_{2,n}, \dots, \alpha_{M,n}\}$, where $\alpha_{m,n} \in \{0, 1\}$. Here, $\alpha_{m,n} = 1$ signifies that source $m$ is scheduled in sampling period $n$, and when $\alpha_{m,n} = 0$, otherwise. The generation indicator is $\beta_{m,n}\in \{0, 1\}$, $\beta_{m,n} = 1$ indicate source $m$ generates a new status update packet at $t_n$. Note that each source only keeps and transmits the freshest packet. Moreover, we adopt the DeepSC\cite{DeepSC} framework to support the textual transmission of the status update.
We will detail the semantic transmission model in the following subsection.

 \begin{figure}[tph]
  \centering
  \includegraphics[width=0.45\textwidth]{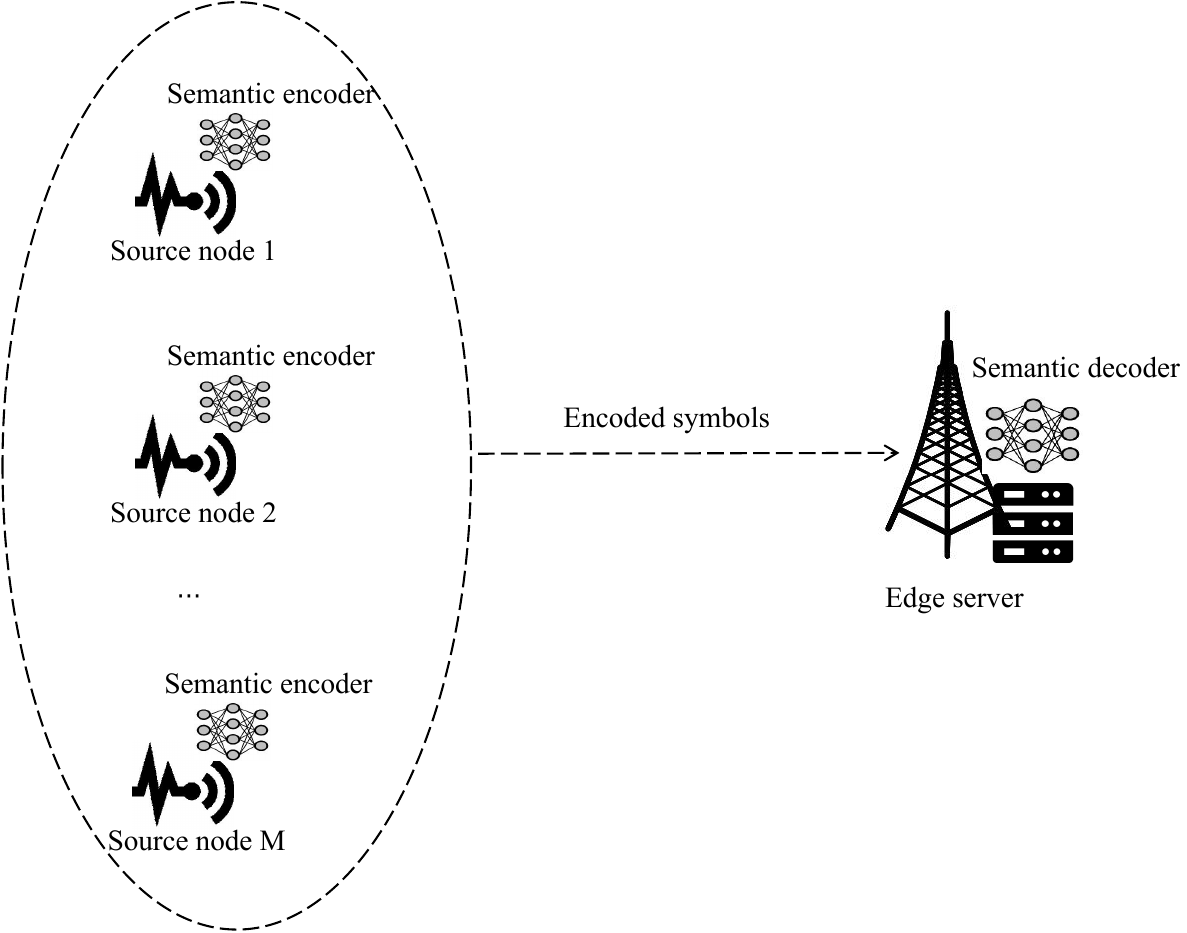}
  \caption{Semantic-aware multi-source status update system.} 
  \label{Fig::sys}
 \end{figure}
\subsection{Semantic transmission model}
Assuming that the wireless channel remains quasi-static within a single sampling period, when source $m$ is scheduled at sampling period $n$, it has to transmit a pilot signal to the edge server, thereby enabling the edge server to acquire the uplink channel parameters $h_{m,n}$. Consequently, the received signal-to-noise ratio (SNR) at the edge server for source $m$ is given by
\begin{align}
    \gamma_{m,n} = \frac{p_m |h_{m,n}|^2}{\sigma^2},
\end{align}
where $p_m$ represents the transmit power of source $m$, $\sigma^2$ is the variance of additive white Gaussian noise (AWGN).

According to \cite{Qin1}, the semantic transmission rate of source $m$ is expressed as
\begin{align}\label{rate}
    R_{m,n} = \frac{W I}{k_{m,n} L} \xi_{m,n},
\end{align}
where $W$ denotes the channel bandwidth, $I$ is the average amount of semantic information contained in each sentence of the packet, $k_{m,n}$ is the number of semantic symbols per word, and $L$ represents the expected number of words per sentence.
Furthermore, $\xi$ is defined as semantic similarity between the transmitted and the reconstructed sentences, $x$ and $\hat{x}$, given by\cite{DeepSC}
\begin{align}
    \xi = \frac{\textbf{B}(x)\textbf{B}(\hat{x})^{\textbf{T}}}{||\textbf{B}(x)||\cdot||\textbf{B}(\hat{x})||},
\end{align}
where $\textbf{B}(\cdot)$ represents the pre-trained BERT model. We also have $0\leq \xi \leq 1$, $\xi = 1$ indicates the highest similarity, and $\xi = 0$ shows no similarity. Therefore, $\xi_{m,n}$ in \eqref{rate} signifies the semantic similarity between the transmitted packet from source $m$ and the recovered packet at sampling period $n$. 


Subsequently, utilizing (\ref{rate}), the transmission latency of source $m$ can be expressed as
\begin{align}
    T_{m,n} = \frac{c_m I}{R_{m,n}},
\end{align}
where $c_m$ is the number of sentences contained in packet $l_{m,n}$. If the transmission latency $T_{m,n}$ exceeds the duration of $\tau$, the transmission is considered unsuccessful.

\section{Age of semantic importance}
\label{sec:3}
In traditional communication system, AoI is a popular metric of information importance by measuring the information time lag at the destination, which can be given as $\Delta^{\text{AoI}}(t) = t - u(t)$, $u(t)$ is the the generation time of the latest received packet. However, AoI deduces information importance solely based on timestamps, without accounting for the semantic content therein. Thus, it is not suitable for semantic communication system. In order to capture the information freshness and the semantic loss in semantic communication system, we propose a novel metric called the Age of Semantic Importance (AoSI). 

Specifically, we measure the semantic importance $\psi$ using the semantic loss caused by missing or error semantic contents\cite{GuoWLS23}. In a text transmission task, it can be defined as

\begin{align}
    \psi =1-\xi = 1 - \frac{\textbf{B}(x)\textbf{B}(\hat{x})^{\textbf{T}}}{||\textbf{B}(x)||\cdot||\textbf{B}(\hat{x})||}.
\end{align}

It is remarkable that the receiver can not measure the semantic importance $\psi$ directly, as it involves both transmitted and received data. Nevertheless, $\psi_{m,n}$, the semantic importance between the source $m$'s transmitted and the recovered packet at sampling period $n$, can be estimated at the receiver. Due to the fact that $\xi_{m,n}$ relies on the neural network structure and channel conditions, we can estimate the semantic similarity $\xi_{m,n}$ as $\tilde{\xi}_{m,n}(k_{m,n}, \gamma_{m,n})$, a function dependent on the number of semantic symbols per word and the received SNR. Therefore, the semantic importance $\psi_{m,n}$ can be estimated as $\tilde{\psi}_{m,n}(k_{m,n}, \gamma_{m,n}) = 1 - \tilde{\xi}_{m,n}(k_{m,n}, \gamma_{m,n})$.

AoSI can be defined as the product of AoI and the semantic importance, given by
\begin{align}\label{AoSI}
    \Delta^{\text{AoSI}}(t) = \Delta^{\text{AoI}}(t) \cdot \psi({u(t)}) = (t - u(t))\cdot \psi({u(t)}).
\end{align}
where $\psi({u(t)})$ is the semantic importance of the latest received packet. Essentially, AoSI serves as a metric encompassing both the timeliness and the semantic importance of the knowledge of a source. In semantic-aware status update systems, the transmitted updates experience semantic loss due to semantic compression and wireless channels, leading to the error in reconstructed status. This semantic loss-induced reconstruction error increasingly penalizes system performance over time. In other words, the less freshness and the lower semantic similarity the received information is, the larger the AoSI is, indicating larger importance and higher priority for scheduling. Note that the proposed AoSI is not a universal metric applicable to all semantic communication scenarios. The definition of semantic importance varies across different scenarios, such as video transmission, image transmission, etc. Therefore, specific definitions of semantic importance should be provided based on the particular scenarios.

\begin{figure} \centering
\subfigure[AoI.] { \label{fig:aoi}
\includegraphics[width=0.45\textwidth]{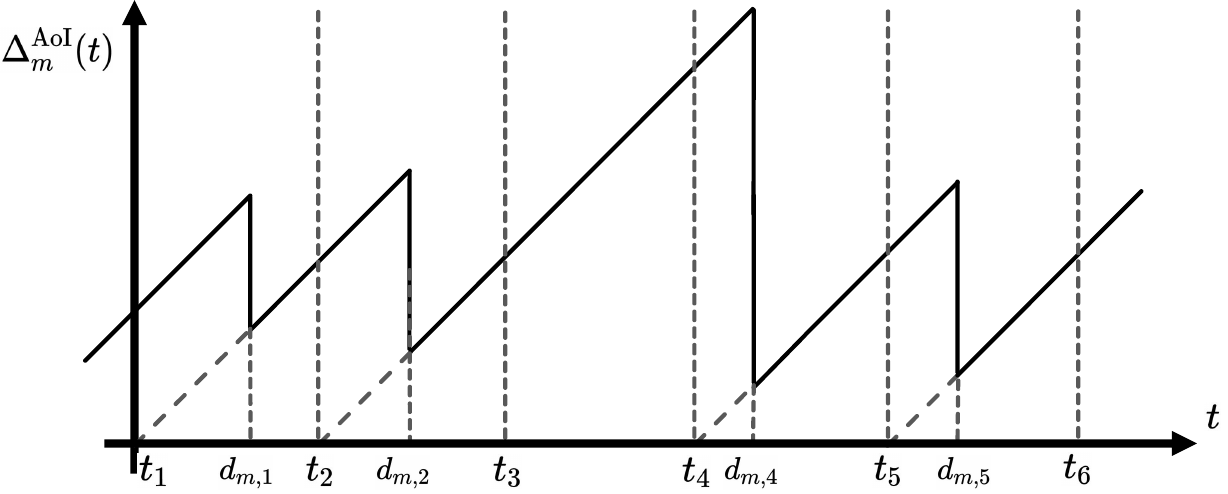}
}
\subfigure[AoSI.] { \label{fig:aosi}
\includegraphics[width=0.45\textwidth]{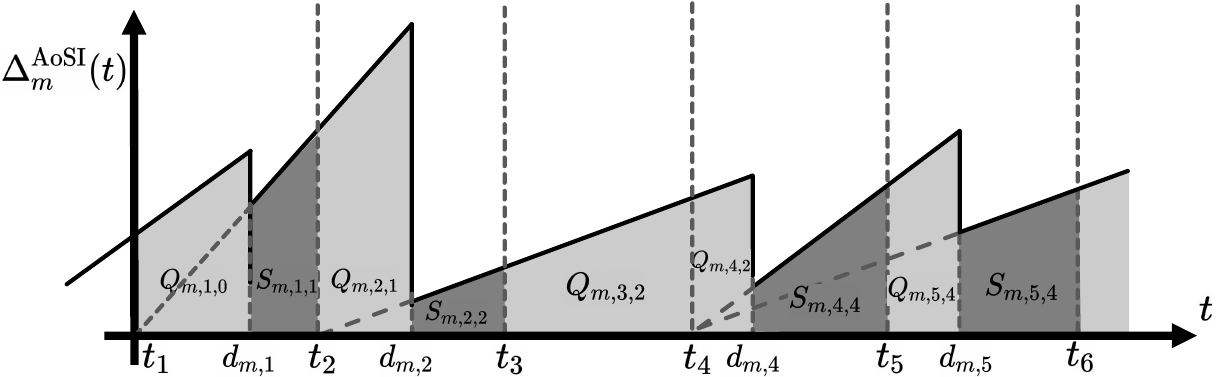}
}
\caption{Evolution of AoI and AoSI. }
\label{fig:AoSI}
\end{figure}

Following the definition of AoSI, we can show the evolution comparison of source $m$'s AoI and AoSI in Fig. \ref{fig:AoSI}, marked as ${\Delta}^{\text{AoI}}_{m}(t)$ and ${\Delta}^{\text{AoSI}}_{m}(t)$. Specifically, the Fig. \ref{fig:aoi} is the AoI evolution of the semantic-aware status update system, while the Fig. \ref{fig:aosi} is the AoSI evolution of the same process. Source $m$ generates update packets at each sampling time instance $t_1, t_2, \dots$, and the successfully transmitted packets are received at $d_{m,1}, d_{m,2}, \dots$. It is remarkable that similar to AoI, the AoSI increases linearly over time, and drops immediately when a new packet is received. The difference is that the slope of AoSI during linear increasing equals to the semantic importance to the last received packet. Hence, the AoSI curve is non-uniform saw-shaped. Moreover, due to limited channel resources and random channel fading, not all the generated packets can be successfully received. In this example, the packet generated at time $t_3$ is not received by the edge server. Furthermore, due to the probabilistic nature of state updates, not all transmitted packets are fresh, some are older packets cached at the source. In this example, the received packet at sampling period $n=5$ is the retransmitted packet generated at time $t_4$ with higher similarity.

Based on the above analysis, we can further give the average AoSI at sampling period $n$ in the considered periodical sampling system as
\begin{align}\label{eq:averageAoSI}
    \bar{\Delta}^{\text{AoSI}}_{m,n} &=\frac{1}{\tau} \int_{t_n}^{t_n + \tau} {\Delta}^{\text{AoSI}}_{m}(t) \, {\rm d}t.
\end{align}

Then, the long term average AoSI can be expressed as
\begin{align}
    \bar{\Delta}^{\text{AoSI}}_{m} &=\lim_{N \to \infty} \frac{1}{NM}\sum_{n=1}^N \sum_{m=1}^M \bar{\Delta}^{\text{AoSI}}_{m,n}.
\end{align}

In order to calculate $\bar{\Delta}^{\text{AoSI}}_{m,n}$, we can firstly compute the area below ${\Delta}^{\text{AoSI}}_{m}(t)$ at sampling period $n$, which corresponds to the integral part in \eqref{eq:averageAoSI}. From Fig. \ref{fig:aosi}, we observe that one sampling period can be divided into two part, the first part $T_{m,n} = d_{m,n} - t_n$ is the transmission time, and the second part $(t_{n+1} - d_{m,n})$ is the waiting time. Therefore, the area below ${\Delta}^{\text{AoSI}}_{m}(t)$ can also be divided in two parts, illustrated as light gray trapezoid and dark gray trapezoid, namely the AoSI caused by transmission and the AoSI caused by waiting, respectively. Specifically, $Q_{m,n,q_{m,n}}$ represents the AoSI caused by transmission of source $m$'s packet during sampling period $n$, with the latest packet at the server side being generated in a sampling period with the index of $q_{m,n}$. Thus, when $T_{m,n}\leq \tau$,  $Q_{m,n,q_{m,n}}$ can be given by
\begin{align}
    Q_{m,n,q_{m,n}} = \frac{1}{2} T_{m,n}\left(\Delta^{\text{AoI}}_m(t_n) + \Delta^{\text{AoI}}_m(d_{m,n})\right)\left(1 - \xi_{m,q_{m,n}}\right).
\end{align}

Similarly, $S_{m,n,b_{m,n}}$ denotes the AoSI caused by waiting during sampling period $n$, where $b_{m,n}$ is the generation index of the newly received packet at sampling period $n$. Thus, when $T_{m,n}\leq \tau$, $S_{m,n,b_{m,n}}$ can be calculated by
\begin{align}    S_{m,n,b_{m,n}} =& \frac{1}{2} \left(\tau - T_{m,n}\right)\nonumber\\
    & \cdot\left(\Delta^{\text{AoI}}_m(d_{m,n}) + \Delta^{\text{AoI}}_m(t_{n+1})\right)\left(1 - \xi_{m,b_{m,n}}\right).
\end{align}

Moreover, when the transmission latency $T_{m,n}$ exceeds the sampling interval $\tau$,  the transmission in this sampling period is considered to be failed. Thus, when $T_{m,n}> \tau$, we have 
\begin{align}   Q_{m,n,q_{m,n}} = \frac{1}{2} \tau \left(\Delta^{\text{AoI}}_m(t_n) + \Delta^{\text{AoI}}_m(d_{m,n})\right)\left(1 - \xi_{m,q_{m,n}}\right).
\end{align}

\begin{align}
S_{m,n,b_{m,n}} = 0.
\end{align}

Therefore, we can give the average AoSI of source $m$ in sampling period $n$ as
\begin{equation}\label{10}
    \bar{\Delta}^{\text{AoSI}}_{m,n} =\frac{1}{\tau}\cdot\begin{cases}

Q_{m,n,q_{m,n}} + S_{m,n,b_{m,n}}, &  \alpha_{m,n} = 1\\
Q_{m,n,q_{m,n}}, & \alpha_{m,n} = 0
    \end{cases}.
\end{equation}

\section{AoSI-aware joint scheduling and resource allocation}
\label{sec:4}

\subsection{Problem formulation}

In order to ensure information freshness and semantic reliability, we formulate the long-term average AoSI minimization problem by jointly optimizing the source scheduling and resource allocation, which can be expressed as
\begin{subequations}
\begin{align}
(\text{P1}):\underset{\boldsymbol{\alpha}_{n}, k_{m,n}}{\min } & \lim_{N \to \infty} \frac{1}{NM}\sum_{n=1}^N \sum_{m=1}^M \bar{\Delta}^{\text{AoSI}}_{m,n} \\
\text { s.t. } & \mathrm{C}1: \alpha_{m,n} \in\{0,1\}, \forall m \in \mathcal{M}, \\
&\mathrm{C}2 :  \sum_{m=1}^M \alpha_{m,n} \leq 1, \\
&\mathrm{C}3 :  k_{m,n} \in\{1,2, \ldots, K\},
\end{align}
\end{subequations}
where $\boldsymbol{\alpha}_{n}$ is the scheduling vector, $k_{m,n}$ is the semantic symbol per word, and $K$ is the upper limit of semantic symbol per word. Specifically, the constraint $\mathrm{C}1$ ensure the binary of scheduling decision ${\alpha}_{m,n}$, the constraint $\mathrm{C}2$ indicates that there is only one source can be scheduled within a sampling period, and $\mathrm{C}3$ defines the range of the number of semantic symbol per word.

Problem $(\text{P}1)$ poses a challenge as a combinatorial optimization problem, which is traditionally addressed through dynamic programming. However, due to its inherent strong coupling and high computational complexity, obtaining an optimal solution for the joint SRA optimization problem remains extremely difficult. 
To solve the optimization problem, we formulate the problem $(\text{P}1)$ as a Markov decision problem (MDP) and develop a DQN-based algorithm to learn a sub-optimal joint SRA policy\cite{DQN}.

\subsection{DQN-based resource allocation}

We deploy a DQN agent at the edge server as a central controller, and model the joint scheduling and resource allocation problem into an MDP. The state $s_{t_n}$, the action $a_{t_n}$, and the reward $r_{t_n}$ of the MDP are as followed,
\subsubsection{State}
The state is the observation of agent at $t_n$, including each source's AoSI $\Delta^{\text{AoSI}}_1(t_n), \Delta^{\text{AoSI}}_2(t_n), \dots, \Delta^{\text{AoSI}}_M(t_n)$, each source's AoI $\Delta^{\text{AoI}}_1(t_n), \Delta^{\text{AoI}}_2(t_n), \dots, \Delta^{\text{AoI}}_M(t_n)$, and the received SNR $\gamma_{1,n}, \gamma_{2,n}, \dots, \gamma_{M,n}$ estimated through pilot signals. The state can be defined as
\begin{align}
    s_{t_n} = \{\Delta^{\text{AoSI}}_1(t_n), \Delta^{\text{AoSI}}_2(t_n), \dots, \Delta^{\text{AoSI}}_M(t_n), \Delta^{\text{AoI}}_1(t_n),\nonumber \\ \Delta^{\text{AoI}}_2(t_n), \dots, \Delta^{\text{AoI}}_M(t_n), \gamma_{1,n}, \gamma_{2,n}, \dots, \gamma_{M,n}\}.
\end{align}

\subsubsection{Action}
The action $a_{t_n}$ determines scheduled source and the number of semantic symbol per word, which can be defined as
\begin{align}
    a_{t_n} = \{\boldsymbol{\alpha}_{n}, k_{m,n}\}.
\end{align}

\subsubsection{Reward}
Since we intend to optimize the average AoSI, the corresponding reward is 
defined as 
\begin{align}
    r_{t_n} = -\frac{1}{M} \sum_{m=1}^M\bar{\Delta}^{\text{AoSI}}_{m,n}.
\end{align}

The goal is to find the optimal SRA policy $\pi^*$, by maximizing the expected cumulative rewards in $\text{Q}$ function, given by
\begin{equation}\label{11}
    \pi^*=\arg \max_{\pi}Q^*(s_{t_n}, a_{t_n}).
\end{equation}

Under such definitions, the DQN framework consists of three parts: evaluation Q-network $Q(s_{t_n}, a_{t_n}; \theta)$, target Q-network $\hat{Q}(s_{t_n}, a_{t_n}; \hat{\theta})$ and a experience replay buffer $E$. At the beginning of each sampling period, sources first send pilot signals to the edge server for the agent to obtain the received SNR $\gamma_{1,n}, \gamma_{2,n}, \dots, \gamma_{M,n}$. Subsequently, the agent uses $\epsilon$-greedy policy to choose actions for the purpose of exploitation and exploration. Following this, the agent store the tuple $(s_{t_n}, a_{t_n}, s_{t_{n+1}}, r_{t_n})$ as training samples in the experience replay buffer. For a certain number of steps, DQN samples a minibatch of training samples and perform training with the assistance of temporal difference (TD) method, the corresponding loss function is defined as follows
\begin{equation}\label{14}
     Loss_{t_n}=(r_{t_n} + \mu \max_{a_{t_{n+1}}} (\hat Q(s_{t_{n+1}}, a_{t_{n+1}}; \hat \theta))-Q(s_{t_n}, a_{t_n}; \theta))^2.
\end{equation}
where $\mu$ is the long term discounted factor. We show the AoSI-aware joint SRA summarized in Algorithm \ref{alg:DQN}.

\begin{algorithm}[tbp]\label{alg:DQN}
    \textbf{Input:} Channel bandwidth $W$, the number of source $M$ \;
    \textbf{Output} {Source scheduling decision $\boldsymbol{\alpha}_n$, semantic symbol per word decision $k_{m,n}$}\;
    \textbf{Initialization:} Initialize the replay buffer $E$, randomly initialize evaluation network with parameter $\theta$ and target network with parameter $\hat \theta$\;
    \For{episode=1,\dots,$E_{\text{max}}$}
    {
    Initialize the state $s_{t_0}$\;
    \For{step=1,\dots,$T_{\text{max}}$}{
    With probability $\epsilon$ select a random action $a_{t_n}$ according to $s_{t_n}$\;
    Execute action $a_{t_n}$ and get the reward $r_{t_n}$ and the state $s_{t_{n+1}}$\;
    Store the transition sample $(s_{t_n}, a_{t_n}, r_{t_n}, s_{t_{n+1}})$ in $E$\;
    Catch a minibatch of replay buffer from $E$\;
    Calculate loss using \eqref{14} and update evaluation Q-network\;
    Every 100 step reset $\hat \theta$ = $\theta$\;}} 
    \caption{AoSI-aware joint SRA}
\end{algorithm}

\section{Simulation results}
\label{sec:5}
This section presents some simulations conducted to evaluate the proposed AoSI-aware joint SRA method. The parameters are set as follows: the channel bandwidth $W$ is 0.1 MHz, each source generates a status update packet with a probability $P_m = 0.8$ per sampling interval, where the status update packet is composed of $c_m = 150$ sentences. The average sentence length $L$ is 20 words, with each word represented by 40 bits. The transmit power $p_m$ is set to 0.1 W, and the variance of AWGN is 0.01. For semantic similarity estimation, a Deep Neural Network (DNN) is utilized to fit the semantic similarity curve as \cite{Qin2}.

Regarding the DQN network structure, both the evaluation network and the target network are configured with one input layer, three hidden layers (64, 256, and 64 neurons), and one output layer. During DQN training, We set 500 sampling intervals for one episode and set the size of replay buffer to 10000. Additionally, the $\epsilon$ value in $\epsilon$-greedy ranges from 0.2 to 0.99, the discount factor $\mu$ is set to 0.9, and the learning rate is $0.001$.

We choose a few separate SRA baseline methods for comparison with the proposed AoSI-aware joint SRA algorithm.
\begin{itemize}
    \item \textbf{Random scheduling:} The Random scheduling method adopts a random policy for source scheduling while utilizing DQN to manage resource allocation.
    \item \textbf{Round-robin:} The Round-robin method utilizes a round-robin source scheduling policy and use DQN for resource allocation.
    \item \textbf{AoI-aware scheduling:} The AoI-aware scheduling method schedules the source with the largest AoI and develop DQN for resource allocation.
    \item \textbf{AoSI-aware scheduling:} The AoSI-aware scheduling method opportunistically selects the source with the highest AoSI at the beginning of one sampling period, and develop DQN to allocate resources to the scheduled source.
\end{itemize}

\begin{figure}[t!]
    \centering
    \includegraphics[width=3.4in]{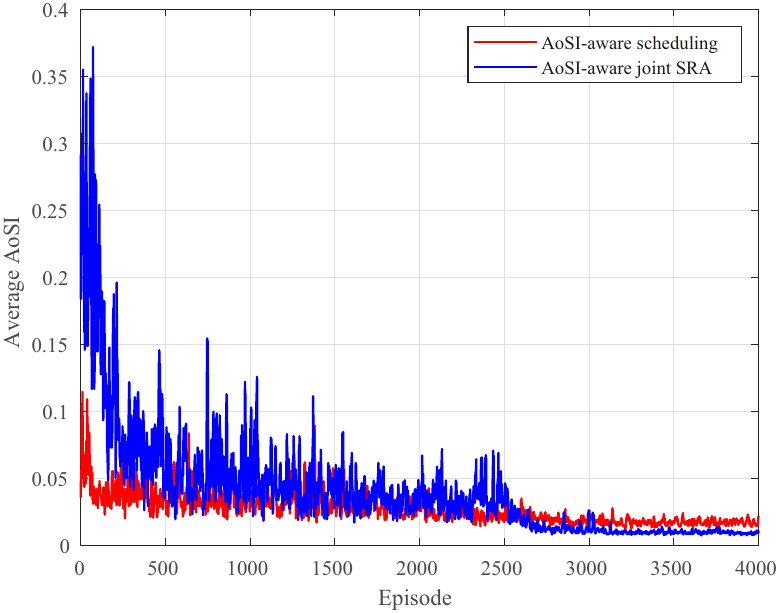}
    \caption{Convergence performance of the proposed AoSI-aware joint SRA.}
    \label{fig:episode}
\end{figure}

Fig. \ref{fig:episode} illustrates the convergence performance of both the proposed AoSI-aware joint SRA method and the separate AoSI-aware scheduling and DQN-based resource allocation method. From this figure, we can find that both curves can converge over training episodes, indicating the effectiveness of the AoSI-aware joint SRA in finding viable scheduling and resource allocation strategies. Moreover, the AoSI-aware joint SRA method has a lower average AoSI and slower convergence speed. This is because the joint optimizing method can simultaneously optimize the source scheduling and resource allocation, at the cost of a larger action space for exploration in the DQN.

\begin{figure}[t!]
    \centering
    \includegraphics[width=3.4in]{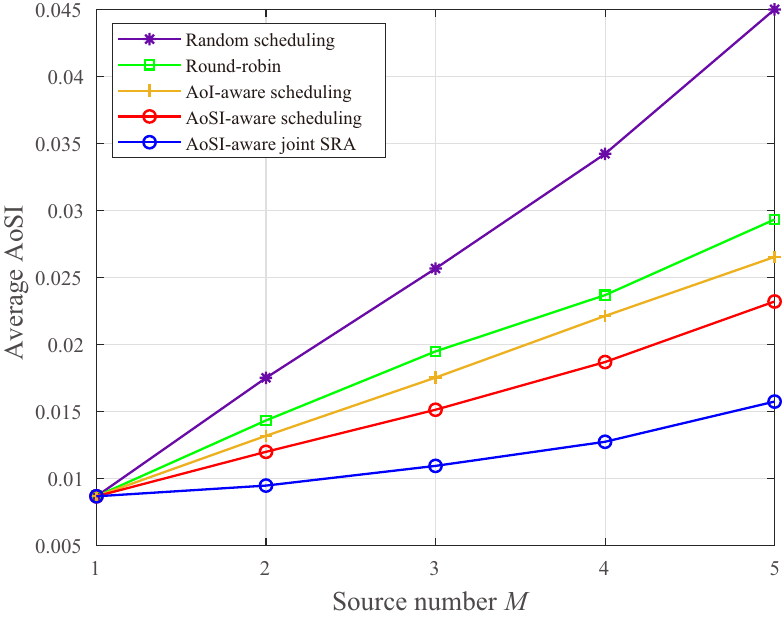}
    \caption{Average AoSI versus different number of sources.}
    \label{fig:user}
\end{figure}

Fig. \ref{fig:user} demonstrates the impact of different source numbers on the average AoSI for several SRA methods, where the number of sources $M\in\{1,2,3,4,5\}$.  It is observed that the average AoSI of all five methods increases with a larger source number, since more sources will cause more intensive competition and lower scheduling frequency. Notably, among the various scheduling and resource allocation methods, the proposed AoSI-aware joint SRA outperforms others, proving its effectiveness.

\begin{figure}[t!]
    \centering
    \includegraphics[width=3.4in]{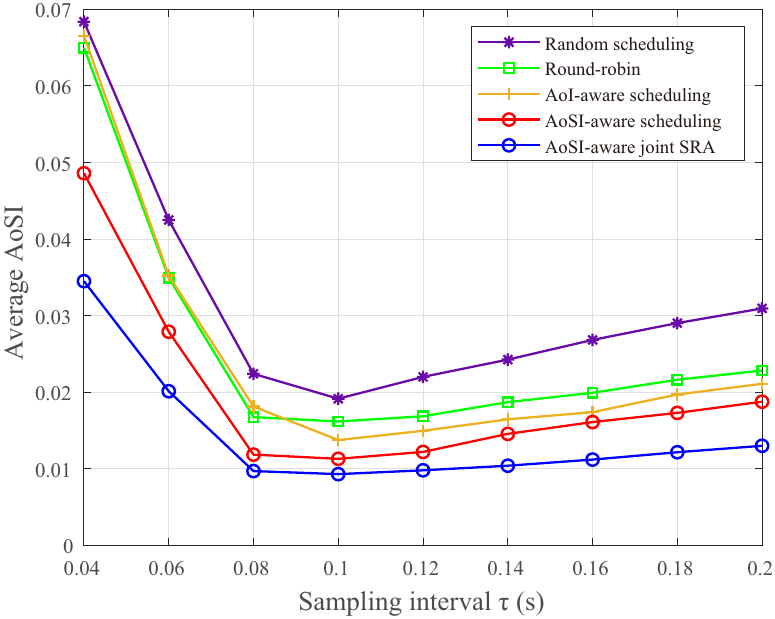}
    \caption{Average AoSI versus different sampling interval.}
    \label{fig:slot}
\end{figure}

Fig. \ref{fig:slot} depicts the impact of different sampling intervals on the AoSI, ranging from 0.05s to 0.25s. We can see that the average AoSI is first decreasing with $\tau$ and then is increasing with $\tau$ when it gets larger, where the optimal sampling interval is around 0.1s. This is because when sampling interval is small, most sources can't finish the communication within $\tau$, or has to choose smaller $k_{m,n}$ which leads to sharp deterioration in semantic similarity. When $\tau$ gets larger, the sampling frequency decrease and therefore deteriorate the average AoSI. It is noteworthy that the proposed AoSI-aware joint SRA still outperforms other methods across varying sampling intervals, providing further evidence of its effectiveness.

\section{Conclusion}
\label{sec:6}
In this paper, we proposed a novel metric called AoSI, which can capture both the information freshness and the semantic similarity in semantic-aware status update system. Through a detailed analysis of AoSI's evolution, we established the system's long-term average AoSI. Then, by formulating the long-term average AoSI optimization problem as an MDP, we proposed a DQN-based joint SRA algorithm to find a sub-optimal solution for source scheduling and the number of semantic symbols. Simulation results served to validate both the introduced AoSI metric and the proposed AoSI-aware joint SRA algorithm, demonstrating that the proposed algorithm outperforms other baseline methods. Moreover, the simulations revealed that the number of sources and the sampling interval in the system both affect the performance of long-term average AoSI.
Future works include the extension to more sampling policies and the investigation of AoSI in more semantic communication scenarios.

\bibliographystyle{IEEEtran}

\bibliography{reference}

\end{document}